\documentclass[aps,prd,showpacs]{revtex4}
\usepackage{graphicx}
\usepackage{amssymb}
\usepackage{amsmath}
\newcommand{\be}{\begin{equation}}
\newcommand{\ee}{\end{equation}}
\newcommand{\ben}{\begin{eqnarray}}
\newcommand{\een}{\end{eqnarray}}
\newcommand{\bes}{\begin{subequations}}
\newcommand{\ees}{\end{subequations}}
\newcommand{\bb}{\bibitem}

\begin{document}
\title{First-order formalism for bent brane}
\author{V.I. Afonso, D. Bazeia, and L. Losano}
\affiliation{Departamento de F\'\i sica, Universidade Federal da Para\'\i ba\\
Caixa Postal 5008, 58051-970 Jo\~ao Pessoa, Para\'\i ba, Brazil}
\date{\today}

\begin{abstract}
This work deals with braneworld scenarios in the presence of real scalar field with standard dynamics. We show that the first-order formalism, which exists in the case of flat brane, can be extended to bent brane, for both de Sitter and anti-de Sitter geometry. We illustrate the results with some examples of current interest to high energy physics.
\end{abstract}
\maketitle

\section{Introduction}
In this work we focus attention on the braneworld scenario described in five-dimensional space-time with warped geometry involving a single extra dimension. The issue is to consider branes in five-dimensional anti-de Sitter (AdS) geometry, embedding four-dimensional AdS, Minkowski (M), or de Sitter (dS) geometry. The scenario was set forward in Refs.~{\cite{rs1,rs2}}, and in the presence of dynamical bulk scalar fields in Ref.~{\cite{gw}}. The five-dimensional warped model of Randall-Sundrum \cite{rs2}, the RS2 model, requires a single infinite extra dimension, and has raised an impressive amount of attention.

Among the important issues, an interesting methodological concern has appeared with the possibility of investigating the equations of motion via first-order differential equations. This issue is addressed for instance in Refs.~{\cite{m,s,f,g,c1,c2,kl,w}}, and it is nicely solved in case the four-dimensional embedded geometry is flat. In the more general case, if the braneworld engenders four-dimensional embedded dS or AdS geometry, some progress was shown in Refs.~{\cite{4,5,w}} and in references therein. 

The present investigation examines the issue of extending the first-order formalism to the case of bent brane, that is, we deal mainly with the possibility of obtaining first-order equations in braneworld scenario driven by scalar field, with embedded geometry of the AdS, M, or dS type. We open a new route to investigate the subject, which we explain below. The main result is related to the recent study of Friedmann-Robertson-Walker (FRW) model for scalar fields coupled to gravity in the four-dimensional cosmological environment \cite{bglm}, with direct interest to dark energy. Evidently, the motivation arises together with the importance of finding first-order formalism, which directly simplifies analysis of the problem and opens new avenues of investigations, both of them contributing to a better understanding of the questions the subject has raised. With due care, our result can be used to generalize the investigation done in Ref.~{\cite{bglm}}. Thus, it is of interest to cosmology, and also to braneworld cosmology \cite{bwc1a,bwc1b,bwc2,bwc3a,bwc3b}. 

Our investigations inspect Einstein's equation and the equation of motion for the scalar field in a very direct way. We consider models described by real scalar fields in five dimensional space-time with AdS geometry, which engenders a single extra dimension and generic four dimensional space-time with AdS, M, or dS geometry. The power of the method that we develop in this Letter is related to an important simplification, which leads to models governed by scalar field potential of very specific form, depending on two functions, $W=W(\phi)$ and $Z=Z(\phi).$ As we show below, we relate the functions $W$ and $Z$ to the warp factor, and this leads to scenarios where the scalar field may also be connected with $W$ and $Z,$ unveiling a new route to investigate the subject. We illustrate our findings with some examples of current interest to high energy physics.
\section{Formalism}
\label{sec:form}

The model that we investigate is described by the action 
\be\label{model}
S=\int\,d^4xdy\;{\sqrt{|g|}\;\left(-\frac14\,R+{\cal L}(\phi,\partial_i\phi)\right)}
\ee
where $\phi$ stands for a real scalar field and we are using ${4\pi G}=1,$ together with dimensionless fields and coordinates. The line element of the five-dimensional space-time is AdS, and can be written as
\be
ds^2_5=g_{ij}dx^idx^j=e^{2A}ds^2_4-dy^2
\ee
for $i,j=0,1,...,4.$ Also, the line element of the four-dimensional space-time can have the form
\be
ds^2_4=dt^2-e^{2\sqrt{\Lambda}t}(dx_1^2+dx_2^2+dx_3^2)
\ee
or
\be
ds^2_4=e^{-2\sqrt{\Lambda}x_3}(dt^2-dx_1^2-dx_2^2)-dx_3^2
\ee
for dS or AdS geometry, respectively. Here $e^{2A}$ is the warp factor and $\Lambda$ represents the cosmological constant of the four-dimensional space-time; the limit $\Lambda\to0$ leads to the line element
\be
ds^2_5=e^{2A}\eta_{\mu\nu}dx^\mu dx^\nu-dy^2
\ee
where $\eta_{\mu\nu},\;\mu,\nu=0,1,2,3,$ describes Minkowski geometry. As usual, the brane is bent for AdS or dS geometry, and flat for Minkowski
space-time. 

The scalar field dynamics is governed by the Lagrange density
\be
{\cal L}=\frac12\,g_{ij}\,\partial^i\phi\,\partial^j\phi-V
\ee
where $V=V(\phi)$ represents the potential, which specifies the model to be considered. The standard scenario is to suppose that both $A$ and $\phi$ are static, depending only on the extra dimension. That is, we set $A=A(y)$ and $\phi=\phi(y),$ and now the equation of motion for the scalar field has the form
\be\label{ephi}
\phi^{\prime\prime}+4A^\prime\phi^\prime=V_\phi
\ee
where prime denotes derivative with respect to $y,$ and $V_\phi=dV/d\phi.$ We first consider Einstein's equation for Minkowski four-dimensional space-ime. This leads to the flat brane case, and we get
\bes\label{ee0}
\ben
A^{\prime\prime}=-\frac23\phi^{\prime2}\label{ee02}
\\
A^{\prime2}=\frac16\phi^{\prime2}-\frac13V(\phi)\label{ee01}
\een
\ees

To get to the first-order formalism, we introduce another function, $W=W(\phi),$ which can be used to see the warp factor as a function of the scalar field. We do this writing the first-order equation
\be\label{A1}
A^\prime=-\frac13 W
\ee
We use this equation and Eq.~(\ref{ee02}) to get to
\be\label{phi1}
\phi^\prime=\frac12W_\phi
\ee
and now the potential in Eq.~(\ref{ee01}) has the form
\be\label{p1}
V=\frac18W^2_\phi-\frac13 W^2
\ee
Evidently, Eqs.~(\ref{A1}) and (\ref{phi1}) solve Eqs.~(\ref{ephi}) and (\ref{ee0}) for the above potential (\ref{p1}). We can also change $W\to-W$
to get another possibility without changing the potential. This result is very interesting, since it simplifies the calculation significantly. As one knows, it was already obtained by other authors in former works \cite{s,f}. See also Refs.~{\cite{bcy,cl,bbn}} for other comments concerning the first-order formalism in arbitrary dimension in supergravity.

We illustrate this case with the example
\be
W_p(\phi)=\frac{2p}{2p-1}\phi^{(2p-1)/p}-\frac{2p}{2p+1}\phi^{(2p+1)/p}
\ee
where $p$ is odd integer. The case $p=1$ is special, and reproduces the $\phi^4$ model in flat spacetime. This model was first introduced in \cite{bmm}, and it was recently considered within the braneworld context in \cite{bfg}. The scalar field is now given by $\phi(y)=\tanh^p(y/p),$ and this gives
\be
A=-\frac13C_+\tanh^{2p}(y/p)+\frac23p\left(C_+-C_-\right)\Bigl\{
\ln[\cosh(y/p)]-\sum_{n=1}^{p-1}\frac{1}{2n}\tanh^{2n}(y/p)\Bigr\}
\ee
with $C_{\pm}=p/(2p\pm1).$

Another interesting example, which leads to analytical investigation of its stability (see below) is given by the superpotential $W=3a\sinh(b\phi).$
It gives the potential
\be
V(\phi)=\frac98 a^2b^2\cosh^2(b\phi)-3a^2\sinh^2(b\phi)
\ee
and the solutions
\bes\ben\label{phiy0}
\phi(y)=\frac{1}{b}{\rm arcsinh}\left[\tan\left(\frac32 ab^2y\right)\right]
\\
\label{ay0}
A(y)=-\frac{2}{3b^2}\ln\left[c\;\sec\left(\frac32 ab^2y\right)\right]
\een\ees
where $a,b,$ and $c$ are real constants. Here we see from $A(y)$ that the metric has a naked singularity at the value $y^*=\pi/3ab^2$. See also Refs.~{\cite{m1,m2,ts,grt}} for other examples of branes supported by scalar fields. 

We now consider the general case of four dimensional AdS, Minkowski or dS geometry. This leads to the bent brane case, and we use Einstein's equation to get
\bes\label{ee}
\ben
A^{\prime\prime}+\Lambda e^{-2A}=-\frac23\phi^{\prime2}\label{ee2}
\\
A^{\prime2}-\Lambda e^{-2A}=\frac16\phi^{\prime2}-\frac13V(\phi)\label{ee1}
\een
\ees
for dS geometry. The case of Minkowski spacetime is obtained in the limit $\Lambda\to0,$ which leads us back to Eqs.~(\ref{ee0}), and
for AdS, we have to change $\Lambda\to-\Lambda.$

The presence of $\Lambda$ makes the problem much harder. Interesting investigations have been already appeared in Refs.~{\cite{f,g,4,5}} and in references therein. Here, however, we follow another route. The key issue is that the presence of $\Lambda$ signals the need of new constraints, and we suggest that
\be\label{A1g}
A^\prime=-\frac13 W-\frac13\Lambda\alpha Z
\ee
and 
\be\label{phi1g}
\phi^\prime=\frac12W_\phi+\frac12\Lambda\beta Z_\phi
\ee
where $Z=Z(\phi)$ is a new and in principle arbitrary function of the scalar field, to respond for the presence of the cosmological constant, and $\alpha$ and $\beta$ are real parameters. These parameters lead to interesting possibilities: if we compare (\ref{A1}) and (\ref{phi1}) with (\ref{A1g}) and (\ref{phi1g}) we see that $\alpha=0$ makes no change in the equation for the warp factor, and $\beta=0$ does not modify the equation for the scalar field. The general extension is inspired in a very recent work \cite{bglm}, in which a first-order formalism was shown to work for cosmology, to describe scalar field in curved space-time. However, the investigation done in Ref.~{\cite{bglm}} corresponds to the case $\alpha=0$ and $\beta\neq0.$ The present procedure is clear, concise, and direct, and can be easily used to provide solutions to specific problems, as we show below.

The general extension will be further examined in a longer work, in preparation. In the present letter, we take $\alpha=1$ and $\beta=1-s,$ which suffices to illustrate the procedure. In this case, in the braneworld scenario generated by scalar field, the potential
$V(\phi)$ is given by
\be
V=\frac18(W_\phi+\Lambda(1-s)Z_\phi)(W_\phi+\Lambda(1+3s)Z_\phi)-\frac13(W+\Lambda Z)^2
\ee
and we now have to include the constraint
\be
W_{\phi\phi}Z_\phi+W_\phi Z_{\phi\phi}+2\Lambda(1-s)Z_\phi Z_{\phi\phi}-\frac43 W Z_\phi-\frac43\Lambda Z Z_\phi=0
\ee
which opens new routes for braneworlds driven by scalar field under the presence of the cosmological constant.

To illustrate the power of this procedure, let us consider the case given by $Z=W.$ This possibility leads to 
\be
\frac32\;c\;W{\phi\phi}-W=0
\ee
where $c={\left(1+(1-s)\Lambda\right)}/{(1+\Lambda)}$. This makes $W$ bounded or not, depending on the sign of c.
We consider an example, described by $W=3a\sinh(b\phi),$ for $b=\pm\sqrt{2/3c},$ and for $c$ positive. In this case we have
\be
V=\frac34 a^2(1+\Lambda)\left[1+(1+3s)\Lambda\right]\cosh^2(b\phi)-3a^2(1+\Lambda)^2\sinh^2(b\phi)
\ee
The scalar field has the form
\be
\phi(y)=\frac1b{\rm arcsinh}[\tan\left(a(1+\Lambda)y\right)]
\ee
and now $A$ is given by
\be\label{ay2}
A(y)=-\frac12\ln[s\,a^2(1+\Lambda)\;{\rm sec}^2\left(a(1+\Lambda)y\right)]
\ee
where we have to take $s\;(1+\Lambda)>0.$ Here we see from $A$ that the metric has a naked singularity at the value $y^*=\pi/2a(1+\Lambda).$ This is similar to the model studied in the second work in Ref.~\cite{g}, which investigates the second-order differential equations of motion. Our investigation is much easier, thanks to the first-order formalism introduced above.

\section{Stability}

The study of stability of the solutions can be done choosing a gauge were the general metric fluctuations have the form
\be
ds^2=e^{2 A(y)}(g_{\mu\nu}+\epsilon\; h_{\mu\nu})dx^\mu dx^\nu-dy^2
\ee
Here  $g_{\mu\nu}=g_{\mu\nu}(x,y)$ represents the four-dimensional AdS, M, or dS metric, $h_{\mu\nu}=h_{\mu\nu}(x,y)$ represents the metric perturbations, and $\epsilon$ is a small parameter. We follow Ref.~{\cite{f}}, introducing the coordinate $z$ to make the metric conformally flat with the choice $dz=e^{-2 A(y)}dy.$ In this case, stability of the solutions leads to the Schr\"odinger equation for metric perturbations, under the choice of transverse and traceless gauge,
\be\label{sch}
-\frac{d^2\psi(z)}{dz^2}+V(z)\psi(z)=k^2\psi(z)
\ee
with
\be
V(z)=-\frac{9\Lambda}{4}+\frac94 A^{\prime 2}(z)+\frac32 A^{\prime\prime}(z)
\ee
for dS geometry; for AdS, we change $\Lambda \rightarrow -\Lambda$, and for Minkowski we take $\Lambda=0$.
For $\Lambda=0$, the Schr\"odinger equation factorizes in the form \cite{f}
\be
\left[-\frac{d}{dz}+\frac34 A^{\prime}(z)\right]\left[\frac{d}{dz}+\frac34 A^{\prime}(z)\right]\psi(z)=k^2\psi(z)
\ee
and so there is no bound-state with negative energy; the zero mode is the zero-energy state $\psi_{0}(z)=e^{-\frac34 A(z)},$ which identifies the ground-state of the quantum mechanical problem.

In the case of $\Lambda \neq 0$, the stability is more involved and should be studied very specifically. For simplicity, however, we firstly examine the case with $\Lambda=0.$ We consider Eq.~(\ref{ay0}), for $b=\sqrt{1/3}$. In this case we have
\be
A(z)=2\ln2-\ln(4c^2+a^2z^2/c^2)
\ee
which leads to
\be
V(z)=12\;\frac{a^2}{c^2}\;\frac{a^2z^2/c^2-c^2}{(a^2z^2/c^2+4c^2)^2}
\ee
This is a volcano-like potential which supports the zero mode and no other bound state. The braneworld scenario is stable, and the graviton is the zero mode which binds to the brane \cite{f,g,c1,c2}. This is the standard scenario in the absence of cosmological constant, and it also appear when we deal with scalar field that engenders unconventional self-interactions \cite{bfg} or with two-field models \cite{bg}.

As the next example, let us consider the same model, by now with $b=\pm \sqrt{2/3.}$ This choice leads to 
\be
A(z)=\ln\left[\frac1c\;{\rm sech}\left(\frac{a}{c}z\right)\right]
\ee
and the potential has the form
\be\label{pflat}
V(z)=\frac94 \frac{a^2}{c^2}-\frac{15}{4}\frac{a^2}{c^2}\;{\rm sech}^2\left(\frac{a}{c} z\right)
\ee
This is a modified P\"oschl-Teller potential. It supports the zero mode and another bound state, massive, with eigenvalue $k^2=2a^2/c^2,$ inside the gap between the zero mode and the continuum, which starts at $k^2=9a^2/4c^2.$ We notice that both the continuum and the massive bound state may be far away from the zero mode for $a>>c.$

We now turn attention to the case of $\Lambda\neq0.$ We consider an interesting example, given by the result obtained in Eq.~(\ref{ay2}), which leads to
\be
A(z)=-\frac12\ln[s\,a^2(1+\Lambda)\;{\cosh}^2\left(q z\right)]
\ee
The potential is now given by
\be\label{pcurved}
V(z)=\frac94(q^2-\Lambda)-\frac{15}{4}q^2{\rm sech}^2(q z)
\ee
where $q^2={(1+\Lambda)/s}$. The model supports two bound states, with the labels $k^2_0=-9\Lambda/4$ and $k^2_1=2q^2-9\Lambda/4.$ For $\Lambda$ negative we get to $AdS_4$ geometry, and the two bound states have positive eigenvalues, making the solution stable. The absence of zero mode indicates that there is no massless graviton, in accord with the Karch-Randall model \cite{kr}; see also \cite{kks,bbg}. On the other hand, we can introduce a massless mode with the restriction $q^2=(1+\Lambda)/s=9\Lambda/8,$ but this leads to a bound state with negative eigenvalue, a tachyonic state with energy $-9\Lambda/4,$ which shows that now $\Lambda$ has to be positive. This result indicates that the possibility of a geometric transition from $AdS_4$ to $dS_4$ would not occur stably. It is interesting to notice that in the above model, the limit $\Lambda\to0$ changes the potential (\ref{pcurved}) to (\ref{pflat}) under the identification $s=c^2/a^2.$ Other interesting comments concerning the above braneworld scenario with non-zero cosmological constant can be found in Ref.~{\cite{g}}, and in references therein.

We remark that the first-order procedure introduced in Sec.~{\ref{sec:form}} has very directly led to the model studied in Ref.~{\cite{g}}, there examined by means of second-order differential equations. The example introduced with $W(\phi)=3a\sinh(b\phi)$ very clearly illustrates the importance of the first-order procedure set forward in the present work.

\section{Ending comments}

In this work we have shown how to write a first-order formalism, to describe scenarios where the braneworld is supported by scalar field, including the possibility of the brane to have AdS, Minkowski, or dS geometry. The crucial ingredient was the introduction of $W=W(\phi)$ and $Z=Z(\phi),$ from which we could express both $A$ and $\phi$ in terms of first-order differential equations, for the potential engendering very specific form. The first-order equations have the general structure ${dA}/{dy}=-(1/3)(W+\Lambda\alpha Z)$ and ${d\phi}/{dy}=(1/2)(W_{\phi}+\Lambda\beta Z_{\phi}),$ where $\Lambda$ is the cosmological constant and $\alpha$ and $\beta$ are real parameters. The importance of the procedure is related not only to the improvement of the process of finding explicit solution, but also to the opening of another route, in which we can very fast and directly investigate the subject.

The approach used in this work focuses on the equations of motion, and provides a direct way to investigate the subject. The procedure used in \cite{f,4} is more sophisticated. In particular, in Ref.~{\cite{4}} there is an interesting investigation based on fake supergravity. The power of the methodology there employed leads to important improvements, concerning both the construction of solutions and the non-perturbative study of stability, valid in arbirtrary dimensions. The procedure of the present study is simpler. It is inspired on a former work on FRW cosmology \cite{bglm}, and offers a concise and direct manner to investigate the problem. Its simplicity poses interesting issues, in particular the investigation of the connection between our approach and the general methodology of Ref.~{\cite{4}}, for instance, the unveiling of plausible relations between the pair of parameters $\alpha$ and $\beta,$ introduced in the first-order equations, with different gauge choices for the complex, $SU(2)$-valued matrix ${\bf W}$ which appears in Sec.~IV of Ref.~{\cite{4}}. Another interesting issue concerns extensions of the present work to models described by two or more scalar fields, which are of direct interest to branes and strings. Although the formal steps may appear straightforward, the simplicity of the
present approach allow practical improvements, concerning the explicit solution of more involved models. These and other related issues
are presently under consideration, and will be the subject of another work.

We would like to thank Francisco Brito for comments and discussions, and CAPES, CLAF/CNPq, CNPq, PADCT/CNPq, and PRONEX/CNPq/FAPESQ for financial support.


\end{document}